\documentclass[aip,apl,amsmath,amssymb,reprint,floatfix]{revtex4-1}

\usepackage{hyperref}
\usepackage{epsfig}
\usepackage{graphicx}
\usepackage{subfigure}
\usepackage{latexsym}
\usepackage{color}
\usepackage{fullpage}
\usepackage{dcolumn}
\usepackage{bm}

\usepackage{ulem}
\usepackage{units}

\definecolor{olivier_comment}{rgb}{0.6,0.6,0.9}

\begin{document}

\title{Magnetism of cobalt nanoclusters on graphene on iridium}

\author{Chi \surname{Vo-Van}}
\affiliation{Institut N\'{e}el, CNRS et Universit\'{e} Joseph Fourier, BP 166, F-38042 Grenoble Cedex 9}

\author{Stefan \surname{Schumacher}}
\affiliation{II. Physikalisches Institut, Universit\"{a}t zu K\"{o}ln, Z\"{u}lpicher Str. 77, D-50937 K\"{o}ln}

\author{Johann \surname{Coraux}}
\email{Johann.Coraux@grenoble.cnrs.fr} \affiliation{Institut N\'{e}el, CNRS et Universit\'{e} Joseph Fourier, BP 166, F-38042 Grenoble Cedex 9}

\author{Violetta \surname{Sessi}}
\affiliation{European Synchrotron Radiation Facility, BP 220, F-38043 Grenoble Cedex}

\author{Olivier \surname{Fruchart}}
\affiliation{Institut N\'{e}el, CNRS et Universit\'{e} Joseph Fourier, BP 166, F-38042 Grenoble Cedex 9}

\author{Nick B. \surname{Brookes}}
\affiliation{European Synchrotron Radiation Facility, BP 220, F-38043 Grenoble Cedex}

\author{Philippe \surname{Ohresser}}
\affiliation{Synchrotron SOLEIL, L'Orme des Merisiers, BP48, Saint-Aubin, F-91192 Gif-sur-Yvette}

\author{Thomas \surname{Michely}}
\affiliation{II. Physikalisches Institut, Universit\"{a}t zu K\"{o}ln, Z\"{u}lpicher Str. 77, D-50937 K\"{o}ln}

\date{\today}

\begin{abstract}
The structure and magnetic properties of Co clusters, comprising from 26 to 2700 atoms, self-organized or not on the graphene/Ir(111) moir\'{e}, were studied \textit{in situ} with the help of scanning tunneling microscopy and X-ray magnetic circular dichroism. Surprisingly the small clusters have almost no magnetic anisotropy. We find indication for a magnetic coupling between the clusters. Experiments have to be performed carefully so as to avoid cluster damage by the soft X-rays. 
\end{abstract}

\maketitle

A noticeable effort is focused at theory level on understanding the magnetic properties of graphene in the presence of transition metals like Fe, Co or Ni. For example, the graphene-mediated exchange interaction between adatoms or impurities each holding a net magnetic moment has been explored and unconventional scaling with distance has been anticipated.\cite{Cheianov2006} Also, magnetic anisotropies as high as required for room-temperature magnetic storage have been predicted for Co dimers.\cite{Xiao2009} So far, experimentalists investigated simpler systems, most prominently the interface between a ferromagnetic layer and graphene. These are indispensable for basic information on proximity-induced magnetic moments in carbon \cite{Weser2010} or the graphene/Co magnetic anisotropy.\cite{VoVan2010} A step towards low-dimensional systems are assemblies of equally sized Fe, Co or Ni clusters comprising a 10-10$^3$ atoms, which are well adapted to the study of the size-dependent magnetic properties. Such clusters can be prepared on epitaxial graphene, \textit{e.g.} on Ir(111),\cite{NDiaye2009} Rh(111),\cite{Sicot2010} or Ru(0001),\cite{Liao2011} using graphene moir\'{e}s as templates.

Using X-ray magnetic circular dichroism (XMCD), we explore the magnetism of Co clusters on graphene/Ir(111), which we relate to the cluster size and distribution by scanning tunneling microscopy (STM). Surprisingly, we find extremely weak magnetic anisotropies, although orbital moments are found to be slightly enhanced compared to bulk values. We also identify degradation of the clusters upon exposure to the X-ray beam.

Magnetic measurements and STM were performed in two interconnected ultra-high vacuum chambers; high resolution STM was performed using the same sample preparation in another system. Ir(111) was cleaned by cycles of Ar$^+$ sputtering and flash annealing to 1500\,K. Graphene on Ir(111) was prepared by chemical vapor deposition of ethene following a two step procedure yielding a closed and perfectly oriented monolayer.\cite{vanGastel2009} Co, Ir, and Pt evaporation was performed close to room temperature. The deposited amount $\theta$ [specified in ML, 1~ML being the surface atomic density of Ir(111)] was calibrated through STM for layers grown directly on Ir(111). XMCD was conducted at the ID08 beamline of the European Synchrotron Radiation Facility, monitored in the total electron yield (TEY) mode using (99$\pm$1\%) circularly polarized light and up to $\pm$5\,T magnetic fields.

\begin{figure}[hbt]
\includegraphics*[width=8cm]{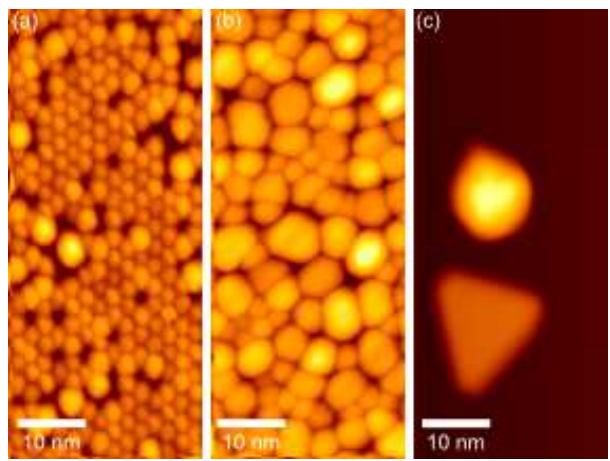}
\caption{STM topographs of metal clusters on graphene/Ir(111): (a) Co$_{26}$ seeded by Pt$_{11}$, (b) Co$_{500}$ seeded by Ir$_{50}$, and (c) pure Co$_{2700}$.} \label{fig:stm}
\end{figure}

Through seeding with Pt and Ir, for small $\theta<$ 1~ML ordered arrays of Co clusters with the moir\'{e} pitch of 2.5\,nm are formed, while for large $\theta>$ 1~ML still dense arrays are obtained, but each cluster spans several moir\'{e} cells and is anchored by several seeds.\cite{NDiaye2009} Deposition of Co without seeds leads to sparse, disordered cluster assemblies. Here we focus on three samples (Tab.~\ref{tab:msml}): triangular lattices of Co$_{26}$ (26 atoms in average) seeded by Pt$_{13}$ (Fig.~\ref{fig:stm}a), Co$_{500}$ seeded by three to four Ir$_{15}$, Ir$_{50}$ hereafter, (Fig.~\ref{fig:stm}b), and pure Co$_{2700}$ (Fig.~\ref{fig:stm}c). Results for Pt$_{13}$Co$_{26}$ are similar to those for Pt$_{13}$Fe$_{26}$ or Ir$_{13}$Co$_{26}$ not detailed here. To test X-ray beam damage effects also small Co$_8$ clusters seeded by Ir$_4$ were probed (sample not listed in Table~\ref{tab:msml}).

\squeezetable
\begin{table}
\caption{\label{tab:msml}Orbital ($m_\mathrm{L}$), spin ($m_\mathrm{S}$) magnetic moments and their ratio ($m_\mathrm{L}$/$m_\mathrm{S}$) at 5~T with the X-ray beam perpendicular to the samples, which were prepared with different deposited amount of seeding and magnetic material ($\theta$), and the average cluster distance ($d$) for each sample (except for Pt$_{13}$Co$_{26}$, where $d$ is the moir\'{e} pitch).}
\begin{ruledtabular}
\begin{tabular}{llllll}
Sample & $\theta$\,(ML) & $d$\,(nm) & $m_\mathrm{S}$\,($\mu_{\mathrm{B}}$) & $m_\mathrm{L}$\,($\mu_{\mathrm{B}}$) & $m_\mathrm{L}$/$m_\mathrm{S}$\\
Pt$_{13}$Co$_{26}$ & 0.13/0.25 & 2.5 & $1.5 \pm 0.2$ & $0.22 \pm 0.02$ & 
0.15$\pm$0.04 \\
Ir$_{50}$Co$_{500}$ & 0.17/1.70 & 4.8 & $1.7 \pm 0.2$ & $0.20 \pm 0.02$& 
0.12$\pm$0.03 \\
Co$_{2700}$ & 0.25 & 30 & $1.7 \pm 0.2$ & $0.18 \pm 0.02$ & 0.11$\pm$0.03 \\
\end{tabular}
\end{ruledtabular}
\end{table}
%% 1 & Pt$_{11}$Co$_{25}$ & 2.5 & 1.5 & 0.22 \\
%% 2 & Ir$_{11}$Co$_{600}$ & 5$\pm$X & 1.72 & 0.2 \\
%% 3 & Co$_{10000}$ & XX$\pm$XX & 1.68 & 0.178 \\
%% 4 & Pt$_{11}$Fe$_{25}$ & 2.04 & 0.268
%% bulk Co 1.62 0.154
%% bulk Fe 2 0.085

\begin{figure}[hbt]
\includegraphics*[width=7cm]{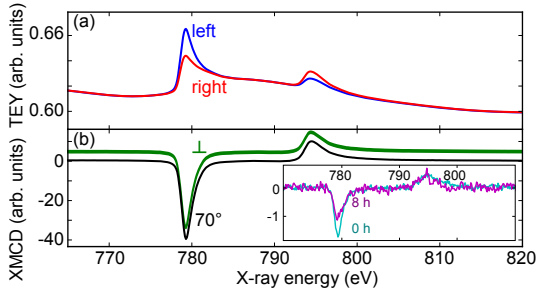}
\caption{\label{fig:XAS-XMCD}(a) TEY from Pt$_{13}$Co$_{26}$ clusters across the L$_{2,3}$ Co absorption edges, for left and right circularly polarized X-rays in perpendicular ($\perp$) incidence, at 5~T and 10~K. (b) XMCD signals for $\perp$ and 70$^\circ$ incidence (vertically shifted for clarity). Inset: XMCD signal normalized to absorption for Ir$_4$Co$_8$, at the beginning and after eight hours of irradiation.} \label{fig2}
\end{figure}

\begin{figure}[hbt]
\includegraphics*[width=6.2cm]{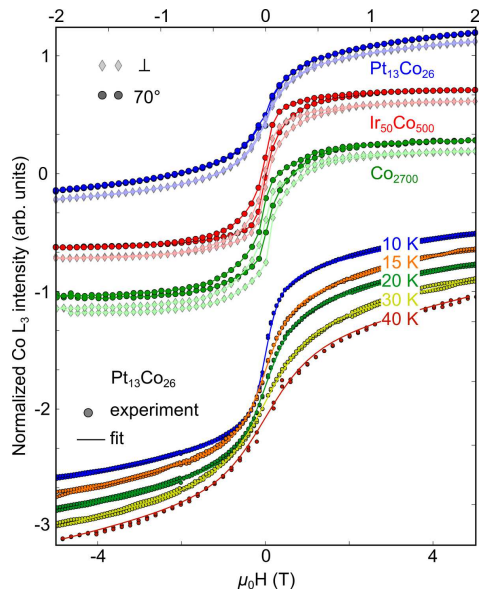}
\caption{\label{fig:HysT}Normalized M-H loops (upper horizontal axis) at 10~K for samples of Table~\ref{tab:msml} for in-plane (70$^\circ$) and perpendicular ($\perp$) incidence. Temperature-dependent M-H loops (lower horizontal axis) and Langevin fits for Pt$_{13}$Co$_{26}$ at $\perp$ incidence. Curves are vertically shifted.} \label{fig3}
\end{figure}

The TEY was measured at 10\,K and $\pm$5\,T across the Co L$_{2,3}$ absorption edges for left- and right circular polarizations of an X-rays beam entering the sample under varying incidence from normal to grazing (70$^\circ$). Subtracting the TEY measured for $\pm$5\,T or opposite polarizations yields the XMCD signal. Figure~\ref{fig:XAS-XMCD}a shows the TEY of Pt$_{13}$Co$_{26}$ at 5\,T in normal incidence for both circular polarizations, while Figure~\ref{fig:XAS-XMCD}b shows the XMCD signal for normal and grazing incidence. No sign of Co oxidation is seen. Magnetization \textit{versus} field (M-H) loops were obtained by subtracting the TEY measured at the Co L$_3$ edge (779\,eV) by the pre-edge value at 774\,eV, and then normalizing to the pre-edge value. We used sum rules\cite{Carra1993} with the number of holes value for bulk and followed the procedure described in Ref.~\onlinecite{Ohresser2001} for deriving the orbital ($m_\mathrm{L}$) and effective spin ($m_\mathrm{S}$, including a dipolar term) magnetic moments at 5\,T, where the M-H loops are close to saturation (Fig.~\ref{fig:HysT}). We find $m_\mathrm{S}=1.5\pm0.2\,\mu_\mathrm{B}$, comparable to the bulk value of 1.62\,$\mu_\mathrm{B}$. On the contrary, $m_\mathrm{L}/m_\mathrm{S}$ is larger than the 0.095 bulk value\cite{Chen1995}. Increasing the cluster size (Ir$_{50}$Co$_{500}$ and Co$_{2700}$) does not substantially modify $m_\mathrm{S}$, but results in a decrease of $m_\mathrm{L}/m_\mathrm{S}$ towards the bulk value (see Tab.~\ref{tab:msml}). Within the limit of uncertainties, $m_\mathrm{L}$ is found identical for all incidences and all sizes.

M-H loops display very weak anisotropies for all cluster sizes (Fig.~\ref{fig:HysT}). For Pt$_{13}$Co$_{26}$ clusters they do not seem to reach saturation and show no hysteresis down to 10\,K (Fig.~\ref{fig:HysT}). A decrease of the zero-field susceptibility is observed as temperature increases (Fig.~\ref{fig:HysT}). Larger clusters have non-zero coercivity, which vanishes at 40$\pm$5\,K.

The enhancement of $m_\mathrm{L}/m_\mathrm{S}$ for small clusters compared to the bulk is typical of small-size objects and arises from the local loss of symmetry (lower coordination or strain).The anisotropy of $m_\mathrm{L}$ (data not shown for grazing incidence X-rays) is here negligible. Within the Bruno model linking the magnetic anisotropy energy~(MAE) with the anisotropy of $m_\mathrm{L}$ (arising from crystal structure, strain and interface hybridization),\cite{Bruno1989} this is consistent with the nearly isotropic hysteresis loops. We do not consider dipolar anisotropy here, as the arrangement of the magnetic atoms in the seeded clusters is unknown. The observation of weak MAE whatever the environment (Pt or Ir seeding) and cluster size is surprising and contrasts with other low-dimensional magnetic systems on metal surfaces.\cite{Ohresser2001,Gambardella2003,Weiss2005}

The absence of coercivity and the decrease of the zero-field susceptibility with increasing temperature are strong indications that the Pt$_{13}$Co$_{26}$ cluster lattices are superparamagnetic, as often the case for nanoclusters. Given the close-to-isotropic magnetic properties of the clusters, the M-H loops were fitted with a Langevin function, \textit{i.e.} assuming no magnetic anisotropy. An additional slope was included in the fits, as discussed latter. The resulting magnetic moment $m=107\pm19$\,$\mu_B$ was found mostly independent of temperature, confirming the relevance of the fitting function. Dividing $m$ by the number of Co atoms per cluster gives 4.1\,$\mu_B$ per atom. This value is unphysical for metallic Co even in low coordination\cite{Gambardella2003} and largely exceeds that derived from sum rules (applied at \unit[5]{T}, \textit{i.e.} beyond the [-1,+1]~T field region where most variations of the Langevin function occur). The occasional coalescence of two or three clusters as visible in Fig.~\ref{fig:stm}a cannot account for the high value of $m$ as only \unit[15]{\%} of the islands are involved. The large $m$ value hence points to correlated spin blocks significantly larger than a single cluster, due to magnetic coupling between neighboring clusters. The origin of this coupling, dipolar, magnetic exchange either direct between Co atoms from neighboring clusters, or indirect through graphene, is at this stage speculative.

Let us also address the linear susceptibility (linear slope), dominating M-H loops above $\pm$2~T (Fig.~\ref{fig:HysT}), which was taken into account in the fitting as an additional slope. The slope is independent from temperature, hinting at an origin different from superparamagnetism. Such linearity up to fields much higher than the expected spontaneous magnetization cannot be ascribed to dipolar energy. It may instead indicate exchange interactions favoring antiparallel or non-collinear magnetization arrangements. Whether this non-collinearity arises within each cluster in a hedge-hog fashion\cite{Respaud1999} or from a block of neighboring clusters remains speculative with the present data.

For larger clusters (Ir$_{50}$Co$_{500}$ and Co$_{2700}$), the vanishing of coercivity (40$\pm$5\,K) is presumably dominated by the blocking temperature of the largest clusters, around 10\,nm in diameter and 3\,nm in height for Co$_{2700}$ assemblies.

Finally, we discuss the degradation of the clusters upon exposure to X-rays. As documented by the inset of Fig.~\ref{fig:XAS-XMCD}b, the XMCD signal decreases through 8\,h irradiation by 40\%, as derived from sum rules. Surfaces measured after the same waiting time, but not exposed to the beam, did not show any reduction of moment. Therefore surface contamination may be excluded. Consequently, the beam induces damage to the sample. Such a degradation is common for fragile magnetic species\cite{Ghigna2001,Lehnert2010} but is usually not observed for metal clusters. We surmise that the X-ray beam promotes the decomposition of graphene by the clusters, similar to what happens at elevated temperature.\cite{Datta2008} Then, carbon enrichment of the clusters might be liable for the reduction of the Co XMCD signal.

%%We note that in the absence of clusters, epitaxial graphene was reported to be stable under soft X-ray exposure.\cite{Zhou2009}

We thank A. Fondacaro for technical assistance. T. M. and S. S. acknowledge financial support from the Deutsche Forschungsgemeinschaft (MI581/17-2), C.V.-V. from the Nanosciences Fondation. Research supported by the French ANR contract ANR-2010-BLAN-1019-NMGEM.

\end{document}